\documentstyle[12pt]{article}
\begin{document}
\thispagestyle{empty}

\begin{center}
\LARGE \tt \bf{Non-Riemannian acoustic spacetime and Magnus field in rotating Bose-Einstein condensates}
\end{center}

\vspace{2.5cm}

\begin{center} {\large L.C. Garcia de Andrade \footnote{Departamento de
F\'{\i}sica Te\'{o}rica - Instituto de F\'{\i}sica Rua S\~{a}o Fco. Xavier 524, Rio de Janeiro, RJ

Maracan\~{a}, CEP:20550-003 , Brasil.

E-Mail.: garcia@dft.if.uerj.br}}
\end{center}

\vspace{2.0cm}

\begin{abstract}
The teleparallel $T_{4}$ version of Riemannian geometry of acoustic spacetime in rotating $2-D$ Bose-Einstein condensates (BEC) is investigated. An experiment is proposed on the basis of phonon ray trajectory around a vortex. The deviation geodesic equation may be expressed in terms of Cartan acoustic torsion. The Riemann curvature is computed in terms of rotation of the fluid. The geodesic deviation equation shows that the acoustic torsion acts locally as a diverging lens and the stream lines on opposite sides of the BEC vortex flow apart from each other. We also show that the Magnus field is cancelled when the acoustic torsion coincides with the rotation of the condensate. This effect is equivalent to the Meissner effect in superconductors. It is interesting to note how the teleparallel acoustic spacetime constrains the physical parameters in the BEC. Here we use the term acoustic torsion since in teleparallelism it is derived from the acoustic metric.
\end{abstract}

\newpage

\section{Introduction}

Recently Peng-Ming Zhang et al \cite{1} have investigated the transverse force on a moving vortex with the acoustic metric.In their framework the vortex appears from the vortex tensor in Lorentz acoustic spacetime in irrotational BEC. They also showed that transverse force on a moving vortex is just the usual Magnus force in irrotational BEC. In this letter we apply the recent idea \cite{2} of non-Riemannian geometry of vortex acoustics in  $^{4}He$ rotating superfluids to a rotating BEC in framework of the teleparallel acoustic spacetime which is the analog model of a theory of gravity that Einstein himself introduce in 1930 \cite{3}, in his discussions with Elie Cartan with the purpose of unifying gravitational and electromagnetic interactions, to extend Peng-Ming Zhang et al results \cite{1} from irrotational to rotating BEC. In particular we show that in the teleparallel case , the Magnus force may vanish due to the presence of acoustic torsion. In our approach here , no use of the topological current theory \cite{4} is needed. Also our approach is not restricted to BEC. Riemannian geometry of vortex acoustics has been investigated in detailed by Fischer and Visser \cite{5,6}. Here we also follow their path to compute the geodesic deviation equation for the phonon ray trajectory around vortex.  This paper is organised as follows: In the section two we present the Riemannian geometry of the model along with the teleparallel counterpart, as well as the proof of the vanishing of the Magnus force, in analogy to the Meissner effect on superconductors. Section 3 presents the computation of the Riemannian geodesic deviation equation and the acoustic torsion lens around the vortex since acoustic torsion in the $T_{4}$ acoustic spacetime is show to be proportional to the vorticity of the fluid. Actually a similar result has been demonstrated in the realm of superfluids recently \cite{2}. Finaly in the conclusions we discuss applications and future prospects. 
\section{Acoustic $T_{4}$ geometry of Rotating BEC}
In this section we shall consider how one can pass from the irrotationa acoustic geometry to a rotational one by simply add an extra term in the definition of the irrotational flow velocity , which is proportional to the vorticity of the fluid. As is well known in classical hydrodynamics this can be expressed by 
\begin{equation}
\vec{v}= \vec{v}^{0}+\vec{\Omega}{\times} \vec{r}
\label{1}
\end{equation}
Here $\vec{v}^{0}={\nabla}{\theta}$ represents the irrotational part of the fluid. The scalar function ${\theta}(t,\vec{r})$ is the phase of the order parameter of the BEC gas, which is given by
\begin{equation}
{\Psi}(\vec{r},t)= \sqrt{\rho} exp(-i\frac{{\theta}m}{\hbar})
\label{2}
\end{equation}
which obeys the nonlinear Schr\"{o}dinger equation ,well known in condensed matter physics as the Gross-Pitaeviskii (GP) equation
\begin{equation}
i\hbar\frac{{\partial}{{\Psi}(\vec{r},t)}}{{\partial}t}=[-\frac{{\hbar}^{2}}{2m}{\nabla}^{2}+ V_{ext}(\vec{r})+{\lambda}|{\Psi}|^{2}]{\Psi}
\label{3}
\end{equation}
where ${\rho}$ is the fluid density. In the Riemannian acoustic geometry \cite{7,8} the geometry is given by the line element of the acoustic spacetime as
\begin{equation}
ds^{2}=\frac{\rho}{c}[-c^{2}dt^{2}+{\delta}_{ab}(dx^{a}-{v^{0}}^{a}dt)(dx^{b}-{v^{0}}^{b}dt)]
\label{4}
\end{equation}
where $\frac{\rho}{c}=constant$, and c is the sound speed in the fluid. Here $(i=0,1,2,3)$ and $(a,b=1,2,3)$. This metric can be transformed simply to the Painleve-Gullstrand metric of acoustic black holes. Applying the coordinate transformation $v^{0}$ going to $v$ into the line element (\ref{4}), we finally obtain the line element of the rotational flow, yet Riemannian, acoustic spacetime 
\begin{equation}
ds^{2}=\frac{\rho}{c}[-c^{2}dt^{2}+{\delta}_{ab}(dx^{a}-({v^{0}}^{a}+{\epsilon}_{abc}{\Omega}^{b}x^{c})dt)(dx^{b}-({v^{0}}^{b}+{\epsilon}_{bde}{\Omega}^{d}x^{e})dt)]
\label{5}
\end{equation}
Note that the last term introduces rotation into the fluid acoustic metric , where ${\Omega}^{b}=(0,{\Omega}^{y},{\Omega}^{z})$ is the rotation vector of the rotating BEC. Bellow we shall see in fact that both components would be identical otherwise, which is imposed by the torsion tensor symmetries. Here ${\epsilon}_{abc}$ is the totally-skew symmetric Levi-Civita symbol. This metric is similar to the Kerr spacetime or the Lense-Thirring one, in the linear approximation of the vorticity. With this metric at hand we now proceed to compute the torsion associated to this metric in the $T_{4}$ framework. Note that when the vorticity of the fluid vanishes that line element reduces to the one corresponding to the irrotational BEC. To easy our task to compute Cartan acoustic torsion we shall consider the Cartan's calculus of exterior differential forms \cite{9} and rewrite down the line element (\ref{5}) in the form	
\begin{equation}
ds^{2}=-({\omega}^{0})^{2}+{\delta}_{ab}({\omega}^{a})({\omega}^{b})
\label{6}
\end{equation}
where the basis one-form ${\omega}^{i}$ are given by
\begin{equation}
{\omega}^{0}=\sqrt{\frac{\rho}{c}}cdt
\label{7}
\end{equation}
\begin{equation}
{\omega}^{a}=\sqrt{\frac{\rho}{c}}(dx^{a}-({{v^{0}}^{a}}+{\epsilon}_{abc}{\Omega}^{b}x^{c})dt)
\label{8}
\end{equation}
By making use of the Cartan equation for torsion
\begin{equation}
T^{i}=d{\omega}^{i}+{{\omega}^{i}}_{k}{\wedge}{\omega}^{k}
\label{9}
\end{equation}
where ${{\omega}^{i}}_{k}$, is the connection one-form which can be made to vanish in the orthonormal frame which from the other Cartan's equation of differential forms calculus
\begin{equation}
{R^{i}}_{jkl}=d{{\omega}^{i}}_{k}+{{\omega}^{i}}_{l}{\wedge}{{\omega}^{l}}_{k}
\label{10}
\end{equation}
yields the vanishing of the Riemann-Cartan tensor ${R^{i}}_{jkl}$, which is exactly the definition of teleparallelism. Thus the components of torsion are the following
\begin{equation}
T^{1}=d{\omega}^{1}= \sqrt{\frac{\rho}{c}}[{\Omega}^{2}-{\partial}_{1}{\partial}_{2}{\theta}] dx^{2}{\wedge}dt 
\label{11}
\end{equation}
\begin{equation}
T^{2}=d{\omega}^{2}= \sqrt{\frac{\rho}{c}}[{\Omega}^{3}-{\partial}_{1}{\partial}_{2}{\theta}] dx^{1}{\wedge}dt 
\label{12}
\end{equation}
\begin{equation}
T^{0}=d{\omega}^{0}= {T^{0}}_{A0}{\omega}^{A}{\wedge}{\omega}^{0}  
\label{13}
\end{equation}
which yields the following components of the acoustic torsion tensor 
\begin{equation}
{T^{2}}_{10}={T^{1}}_{20}=\sqrt{\frac{\rho}{c}}[{\Omega}-{\partial}_{1}{\partial}_{2}{\theta}] 
\label{14}
\end{equation}
\begin{equation}
{T^{0}}_{A0}=\sqrt{\frac{\rho}{c}}{\partial}_{A}{\theta}
\label{15}
\end{equation}
where here $(A=1,2)$. Since by symmetries of the torsion tensor ,${T^{i}}_{jk}={T^{i}}_{[jk]}$ and ${T_{ijk}}={T_{(ij)k}}$ expressions (\ref{15}) vanish, one realizes that at least locally teleparallelism imposes the constrain on the speed of sound in BEC to be $c=c(t)$. The Magnus transverse force on the vortex could be computed from the autoparallel equation
\begin{equation}
\frac{d^{2}{x^{i}}}{ds^{2}}= -[{{\Gamma}^{i}}_{jk}-{T^{i}}_{(jk)}]\frac{dx^{j}}{ds}\frac{dx^{k}}{ds}
\label{16}
\end{equation}
In order that the Magnus force vanishes , from the autoparallel equation
\begin{equation}
\frac{d^{2}{x^{1}}}{ds^{2}}= -\sqrt{\frac{\rho}{c}}[{\Omega}-{\partial}_{1}{\partial}_{2}{\theta}]\frac{dx^{2}}{ds}\frac{dx^{0}}{ds}
\label{17}
\end{equation}
one has to obey the the following constraints for the values of acoustic torsion 
\begin{equation}
{\partial}_{1}{\partial}_{2}{\theta}=\frac{1}{2}{\Omega}
\label{18}
\end{equation}
\begin{equation}
{T_{20}}^{1}= \frac{1}{2}\sqrt{\frac{\rho}{c}}{\Omega}
\label{19}
\end{equation}
which represents the result already obtained for the case of superfluids in reference \cite{2}. Thus note that the acoustic torsion is a cachteristic of the roation of the condensate. Besides here we call the acoustic torsion since it is in teleparallel acoustic spacetime derived from the acoustic metric.
\section{Riemannian curvature in BEC and gravitational analog of diverging lens}
In this section we shall compute the curvature quantities associated with the rotating BEC. We also compute the geodesic deviation equation and show that the curvature phonon sees, acts locally as a diverging lens with a complex curvature frequency. Let us now consider the Riemann curvature components by applying the  Gauss-Codazzi decomposition used by Fisher and Visser in reference \cite{10}. Applying their expressions to our particular example one obtains, first the Riemann-Christoffel connections
\begin{equation}
{{\Gamma}^{t}}_{ab}=D_{ab}=({\partial}_{a}v_{b}+{\partial}_{b}v_{a})= ({\partial}_{a}{\partial}_{b}+{\partial}_{b}{\partial}_{a}){\theta} 
\label{20}
\end{equation} 
note that the rotation of the fluid does not appear in this connection component.
The remaining connection components are 
\begin{equation}
{{\Gamma}^{t}}_{tt}= v_{a}v_{b}D_{ab}=\frac{1}{2}({\nabla}{\theta}+\vec{\Omega}{\times}\vec{r}).{\nabla}(\vec{v_{0}}^{2}+2\vec{v_{0}}.{\Omega}{\times}\vec{r})
\label{21}
\end{equation} 
where we have dropped the second order terms in the vorticity, and 
\begin{equation}
{{\Gamma}^{t}}_{at}=-v_{b}D_{ab}=-({\partial}_{b}{\theta})({\partial}_{a}{\partial}_{b}+{\partial}_{b}{\partial}_{a}){\theta} 
\label{22}
\end{equation} 
\begin{equation}
{{\Gamma}^{a}}_{bc}= v_{a}D_{bc}= ({\partial}_{c}{\theta})({\partial}_{a}{\partial}_{b}+{\partial}_{b}{\partial}_{a}){\theta} 
\label{23}
\end{equation}
\begin{equation}
{{\Gamma}^{a}}_{tt} = -{\partial}_{t}{\partial}_{a}{\theta}-\frac{1}{2}({\delta}_{ab}-v^{a}v^{b}){\partial}_{b}({\vec{v^{0}}}^{2}+2{\Omega}{\times}\vec{r}.\vec{v^{0}})
\label{24}
\end{equation}
where the first term vanishes since $U^{i}$ does not depend on time  
\begin{equation}
{{\Gamma}^{a}}_{tb} = -v^{a}v^{c}D_{bc}
\label{25}
\end{equation}
\begin{equation}
R_{abcd} =(D_{cd}D_{ab}-D_{bd}D_{ac})={R^{0}}_{abcd}
\label{26}
\end{equation}
\begin{equation}
R_{tabc} = v_{d}(D_{cd}D_{ab}-D_{bd}D_{ac})
\label{27}
\end{equation}
\begin{equation}
R_{tatb} = -(D^{2})_{ab}- v_{c}D_{c,ab}+ v_{c}v_{d}(D_{cd}D_{ab}-D_{bc}D_{ad})
\label{28}
\end{equation}
 To simplify the computations we consider the Riemann components in a orthonormal tetrad Minkowski frame $(e^{a}_{\mu})$, where ${\mu}=0,1,2,3$. The metric in the tetrad frame is ${\eta}_{ab}$ where $(a=0,1,2,3)$, is given in terms of coordinate Riemann metric as   
\begin{equation}
g_{{\mu}{\nu}} = {\eta}_{ab}{e^{a}}_{\mu}{e^{b}}_{\nu}
\label{29}
\end{equation}
where we use the same simple gauge used by Fischer and Visser \cite{9} as ${e^{t'}}_{t}=-1$, ${e^{i'}}_{t}=-v^{i}$ and ${e^{j'}}_{i}= {{\delta}^{j'}}_{i}$. Finally the Riemann components in this gauge are
\begin{equation}
R_{t'y't'y'} = -D_{yx}D_{xy}=-\frac{c}{\rho}{T_{20}}^{1}=-\frac{1}{2}\sqrt{\frac{c}{\rho}}{\Omega}
\label{30}
\end{equation}
\begin{equation}
R_{x'y'x'y'} = -{D^{2}}_{yx}
\label{31}
\end{equation}
With these curvature components we are now able to calculate the geodesic deviation equation
\begin{equation}
\frac{D^{2}}{d{\lambda}^{2}}[{\delta}y]+[(R_{y't'y't'}+R_{y'x'y'x'})(u^{t'})^{2}][{\delta}y]=0
\label{32}
\end{equation}   
where ${\delta}y$ is a space-like separation of a family of geodesics in the $x-direction$ with tangent vector ${\vec{u^{a}}}=(u^{t'},u^{t'},0,0)$ and $\vec{n}=(0,0,{\delta}y,0)$ is the separation vector of geodesic deviation. Substitution of the curvature expressions (\ref{30}) and (\ref{27}) into equation (\ref{32}) yields
\begin{equation}
\frac{D^{2}}{d{\lambda}^{2}}[{\delta}y]-\sqrt{\frac{c}{\rho}}{\Omega}(u^{t'})^{2}[{\delta}y]=0
\label{33}
\end{equation}
Since the geodesic deviation equation can be considered as a parametrically driven harmonic oscillator \cite{9} in the affine parameter ${\lambda}$ with frequency   
\begin{equation}
{\omega}({\lambda})= u^{t'}\sqrt{R_{y't'y't'}+R_{y'x'y'x'}}
\label{34}
\end{equation}
in the turbulent shear layer considered here the frequency is then imaginary or ${\omega}({\lambda})= i\sqrt{\frac{c}{\rho}}{\Omega}u^{t'}$ which leads us to the physical conclusion that the Riemann curvature acts locally as a diverging lens in the teleparallel effective spacetime of the rotating BEC.     

\section{conclusions}
A  teleparallel framework for the rotating BEC is displayed. The Magnus force is shown to vanish if the acoustic torsion coincides with the rotation of the BEC. Cartan torsion vanishes in this case when rotation also vanishes. The non-Riemannian geometry of rotating BEC imposes several constraints on possible experiment to test the model. We could also say that due to the proportionality between torsion and rotation the vortex formation would be favoured by the presence of torsion as happens in the case discussed by Kawaguchi et al \cite{11} in the topological vortex formation in BEC under the action of a Newtonian gravitational field; the basic difference being that in their case BEC is not rotating.
\section*{Acknowledgement}
I thank  UERJ for financial support. 

\end{document}